\documentclass[a4paper,11pt]{article}
\usepackage{pos}
\usepackage{graphicx}
\usepackage{dcolumn}
\usepackage{bm}
\usepackage{amsmath}
\usepackage{braket}
\usepackage{slashed}
\usepackage{epstopdf}
\usepackage{placeins}
\usepackage{multirow}
\usepackage{makecell}
\usepackage{soul}
\usepackage{braket}
\usepackage[shortlabels]{enumitem}
\usepackage{placeins}

\usepackage[font=small, labelfont=bf, textfont={small,it}]{caption}

\newcommand{\beq}{\begin{eqnarray}}
	\newcommand{\eeq}{\end{eqnarray}}
\newcommand{\beqnn}{\begin{eqnarray*}}
	\newcommand{\eeqnn}{\end{eqnarray*}}

\newcommand{\Tr}{\ensuremath{\mathrm{Tr}}}

\newcommand{\stag}{\ensuremath{\mathrm{stag}}}

\newcommand{\R}{\ensuremath{\mathrm{R}}}
\newcommand{\phys}{(\ensuremath{\mathrm{phys}})}

\title{Chiral condensate from the spectrum of the staggered Dirac operator}

\author[a]{Claudio Bonanno}
\author*[b]{Francesco D'Angelo}
\author[b]{Massimo D'Elia}

\affiliation[a]{Instituto de F\'isica Te\'orica UAM-CSIC, c/ Nicol\'as Cabrera 13-15, Universidad Aut\'onoma de Madrid, Cantoblanco, E-28049 Madrid, Spain}

\affiliation[b]{Università di Pisa and INFN Sezione di Pisa, Largo B.~Pontecorvo 3, I-56127 Pisa, Italy}

\emailAdd{claudio.bonanno@csic.es}
\emailAdd{francesco.dangelo@phd.unipi.it}
\emailAdd{massimo.delia@unipi.it}

\abstract{The chiral condensate is computed from the mode number of the staggered Dirac operator. This result is compared with those obtained with other approaches, based on the quark mass dependence of the topological susceptibility and of the pion mass.}

\FullConference{The 40th International Symposium on Lattice Field Theory (Lattice 2023)\\
July 31st - August 4th, 2023\\
Fermi National Accelerator Laboratory\\}

\begin{document}
\maketitle

\section{Introduction}

Chiral Perturbation Theory (ChPT)~\cite{Gasser:1983yg,Gasser:1984gg} is a very useful tool in describing the QCD low-energy dynamics and the role that flavour symmetries play in it. In order to fully specify the SU(2) ChPT at leading order, two low energy constants (LECs) are needed: the \textit{chiral condensate} $\Sigma$ and the \textit{pion decay constant} $F_\pi$, defined as follows:
\beq\label{eq:Sigma_def}
\Sigma \equiv - \lim_{m_u,m_d\to 0} \braket{\overline{u} u},
\eeq
\beq\label{eq:Fpi_def}
F_\pi \equiv \lim_{m_u,m_d\to 0} \frac{1}{M_\pi}\bra{\Omega} \overline{u}\gamma_4\gamma_5 d \ket{\pi(\vec{p}=\vec{0})}.
\eeq
There are two different approaches to determine these LECs: the first one consists in comparing the effective theory predictions with experimental data (such as hadron masses); the second one is more theoretically driven, and is based on the matching done with quantities computed in the full theory by means, for instance, of lattice QCD simulations.

The determination of the chiral condensate from lattice QCD has been extensively pursued in the literature~\cite{ETM:2009ztk,Cichy:2013gja,Brandt:2013dua,Engel:2014eea,Bazavov:2010yq,Borsanyi:2012zv,Budapest-Marseille-Wuppertal:2013vij,Boyle:2015exm,Cossu:2016eqs,Aoki:2017paw,Bonanno:2023ypf}. In this talk, we present a novel computation of the flavour SU(2) chiral condensate in $N_f=2+1$ QCD with staggered fermions, by using the mode number approach~\cite{Giusti:2008vb} based on the Banks--Casher relation~\cite{Banks:1979yr}. We also check our results with methods based on the relationship of $\Sigma$ with the pion mass $M_\pi$ and the topological susceptibility $\chi$. In this proceedings we will summarize the main findings of Ref.~\cite{Bonanno:2023xkg}, to which we refer the reader for more details.

\section{Lattice setup and determination of the Lines of Constant Physics (LCPs)}

We perform simulations of $N_f=2+1$ QCD on different lines of constant physics characterized by different values of the pion masses. We use the tree-level Symanzik improved Wilson action for the gauge sector and the rooted stout staggered discretization for the fermionic determinant (see Ref.~\cite{Bonanno:2023xkg} for the simulation parameters).

According to Eq.~(\ref{eq:Sigma_def}), the flavour SU(2) chiral condensate is defined in the limit $m_l\to0$ so we want to keep the renormalized strange quark mass fixed between the different LCPs. In the following, we explain how these LCPs are constructed. To fix the notation, we call $\beta$, $m_u=m_d\equiv m_l$ and $m_s$ the bare parameters of the simulation. The starting point is the physical LCP~\cite{Aoki:2009sc, Borsanyi:2010cj, Borsanyi:2013bia}:
\beq\label{eq:physical_LCP}
\left( \beta^{\phys}, m_l^{\phys}, m_s^{\phys} \right),
\eeq
corresponding to $M_\pi^{\phys}\simeq135~\mathrm{MeV}$ and $R^{\phys}\equiv m_l/m_s\simeq 1/28.15$. The other LCPs are obtained from the physical one simply by increasing $m_l$ while keeping $\beta$ and $m_s$ fixed. 

When doing this, one has to pay attention not to change too much the scale: indeed, according to arguments based on the loop expansion of the fermionic determinant, when increasing $m_l$, one has an effective $\beta_{\mathrm{eff}}<\beta$ and a lattice spacing larger than the one corresponding to the physical LCP. This is the reason why we checked, by setting the scale with the $w_0$ parameter based on gradient flow~\cite{BMW:2012hcm}, that any variation of the lattice spacing is within the error bars (of order $\sim2\%$) and this justifies the claim that, at our level of precision, the renormalized strange quark mass can be considered fixed.

By using the strategy previously described, we determined three LCPs with $m_l=4,6,9~m_l^{\phys}$. The ensembles at the physical point are the same used in Ref.~\cite{Athenodorou:2022aay}.

\section{$\Sigma$ from the mode number}

In this Section, the computation of $\Sigma$ with the mode number approach is presented. The low-lying part of the spectrum of the Dirac operator contains all the information about the chiral condensate via the Banks--Casher relation:
\beq\label{eq:banks_casher}
\lim_{\lambda\to 0}\lim_{m\rightarrow0}\lim_{V\rightarrow\infty}\rho(\lambda,m) = \frac{\Sigma}{\pi},
\eeq
where $\rho(\lambda,m)$ is the spectral density of the eigenvalues $i\lambda$ of $\slashed{D}$,
\beq\label{eq:spectral_density_def}
\rho(\lambda,m)=\frac{1}{V}\sum_{k} \langle \delta(\lambda-\lambda_k)\rangle
\eeq
and $V$ is the 4-dimensional space-time volume.

On the lattice, a more convenient quantity to work with is the \textit{mode number}, i.e., the integral of the spectral density up to a certain threshold mass $M$:
\beq\label{eq:mode_number}
\langle\nu(M)\rangle = V \int_{-M}^{M} \rho(\lambda,m)~d\lambda.
\eeq
This quantity contains exactly the same amount of information of the spectral density, and if one is sufficiently close to the origin, one can use the Banks--Casher relation to obtain:
\beq\label{eq:banks_casher_mode_number}
\langle\nu(M)\rangle = \frac{2}{\pi}V\Sigma M+o(M).
\eeq
The mode number approach consists in evaluating the behaviour of $\langle\nu(M)\rangle$: if one is able to find a linearity region sufficiently near the origin (in order to neglect higher order corrections in $M$), then it is possible to perform a linear fit and extract the chiral condensate according to Eq.~(\ref{eq:banks_casher_mode_number}).

When using staggered fermions, it is necessary to take into account the well-known taste degeneration: indeed, in order to obtain the physical mode number $\langle\nu(M)\rangle$, one has to divide the staggered counterpart $\langle\nu_\stag(M)\rangle$ by the number of different tastes $n_t=2^{d/2}$ ($n_t=4$ in 4 dimensions). Finally, since $\langle\nu (M)\rangle$, $M/m_s$ and $\Sigma m_s$ are RG-invariant quantities in the staggered formulation, Eq.~(\ref{eq:banks_casher_mode_number}) can be written in the following manifestly-RG-invariant way:
\beq\label{eq:banks_casher_RG_inv}
\langle \nu(M/m_s) \rangle = \frac{\langle \nu_\stag (M/m_s)\rangle}{4}=\frac{2}{\pi}V[\Sigma m_s] \left ( \frac{M}{m_s}\right ) + o\left (\frac{M}{m_s}\right ).
\eeq
In this case, from the linear behaviour one is able to extract $\Sigma m_s$. In order to recover $\Sigma_\mathrm{R}$, we use the value $m_s=92.4(1.5)~\mathrm{MeV}$ in the $\overline{\mathrm{MS}}$ renormalization scheme at $\mu=2~\mathrm{GeV}$, obtained in Ref.~\cite{Davies:2009ih} with lattice simulations of $N_f=2+1$ QCD with staggered fermions.

\begin{figure}[!htb]
	\centering
	\includegraphics[scale=0.37]{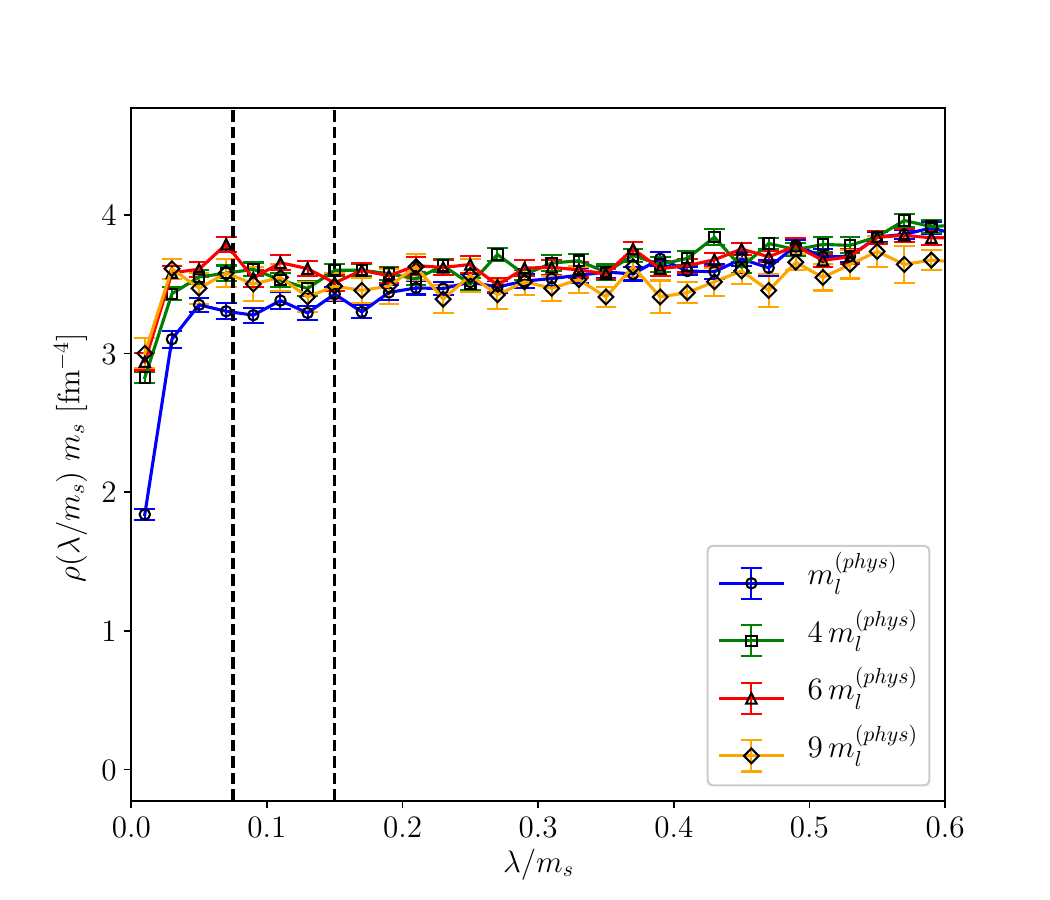}	
	\includegraphics[scale=0.35]{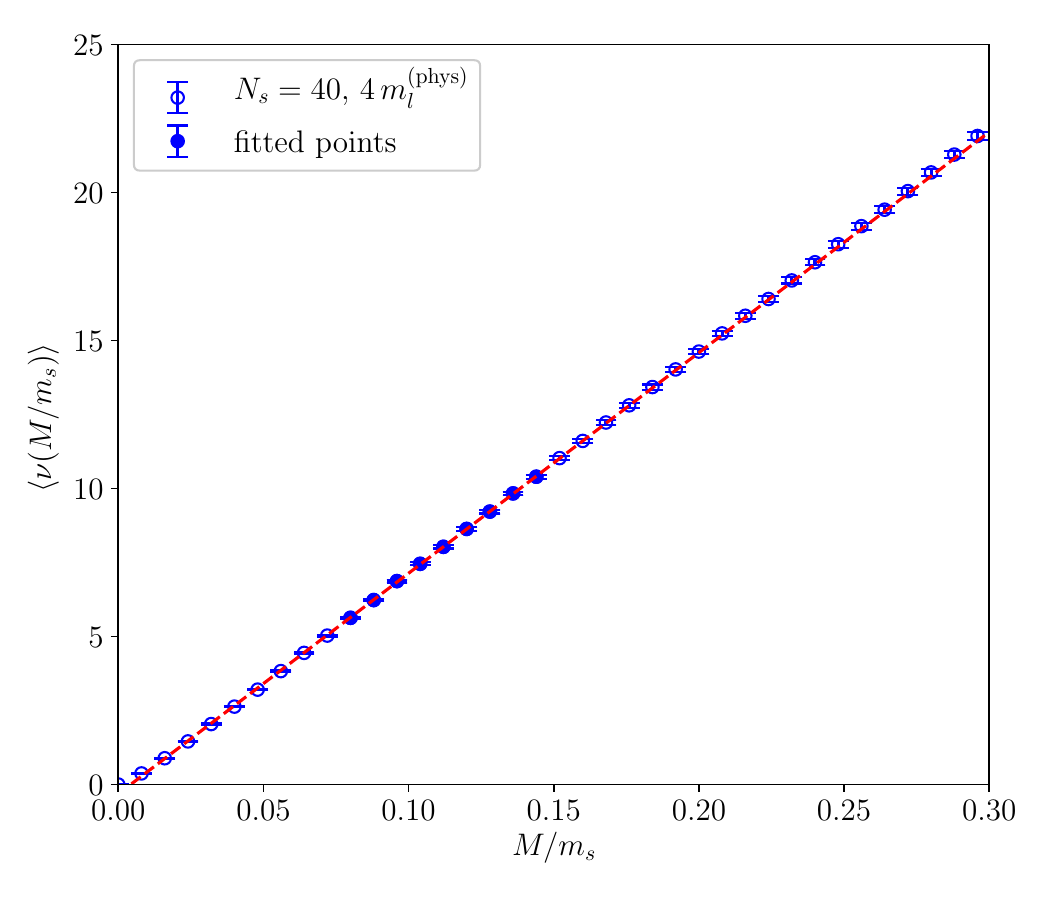}
	\caption{Left: behaviour of the RG-invariant spectral density $m_s\rho(\lambda/m_s)$ as a function of the RG-invariant ratio $\lambda/m_s$ for the lattices with the finest lattice spacing, and for all explored values of $m_l$. The range between the two dashed vertical lines, $\lambda/m_s\in[0.075,0.15]$, is the one chosen to perform the linear best fit of the mode number. Right: linear best fit of the physical mode number $\braket{\nu} = \braket{\nu_\stag}/4$ as a function of $M/m_s$ for the finest lattice spacing for $m_l=4~m_l^{\phys}$. Filled points are those included in the best fit, depicted as a dashed line. Plots taken from~\cite{Bonanno:2023xkg}.}
	\label{fig:spectral_density}
\end{figure}

In order to find the linearity region in the mode number, the idea is to look at the normalized spectral density $m_s\rho(\lambda/m_s)$ to find a common plateau for all the ensembles used. Being the renormalized strange quark mass kept fixed between the different LCPs, it is possible to use the same renormalized fit range to properly compute the chiral condensate. According to the left hand side plot of Fig.~\ref{fig:spectral_density}, we choose as fit range $M/m_s\in[0.075,0.15]$. On the right hand side, an example of extraction of the effective condensate $\Sigma(a, R)$ from the linear fit is shown.

\begin{figure}[!htb]
	\centering
	\includegraphics[scale=0.37]{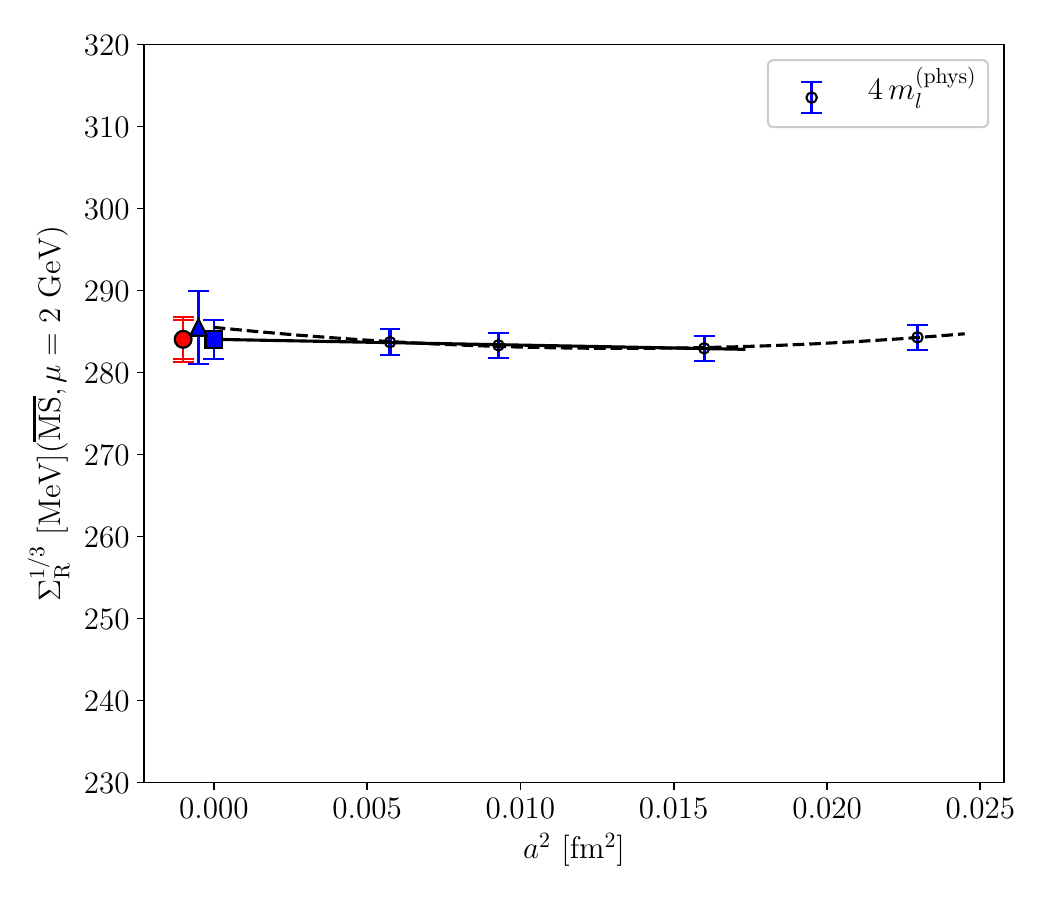}
	\includegraphics[scale=0.37]{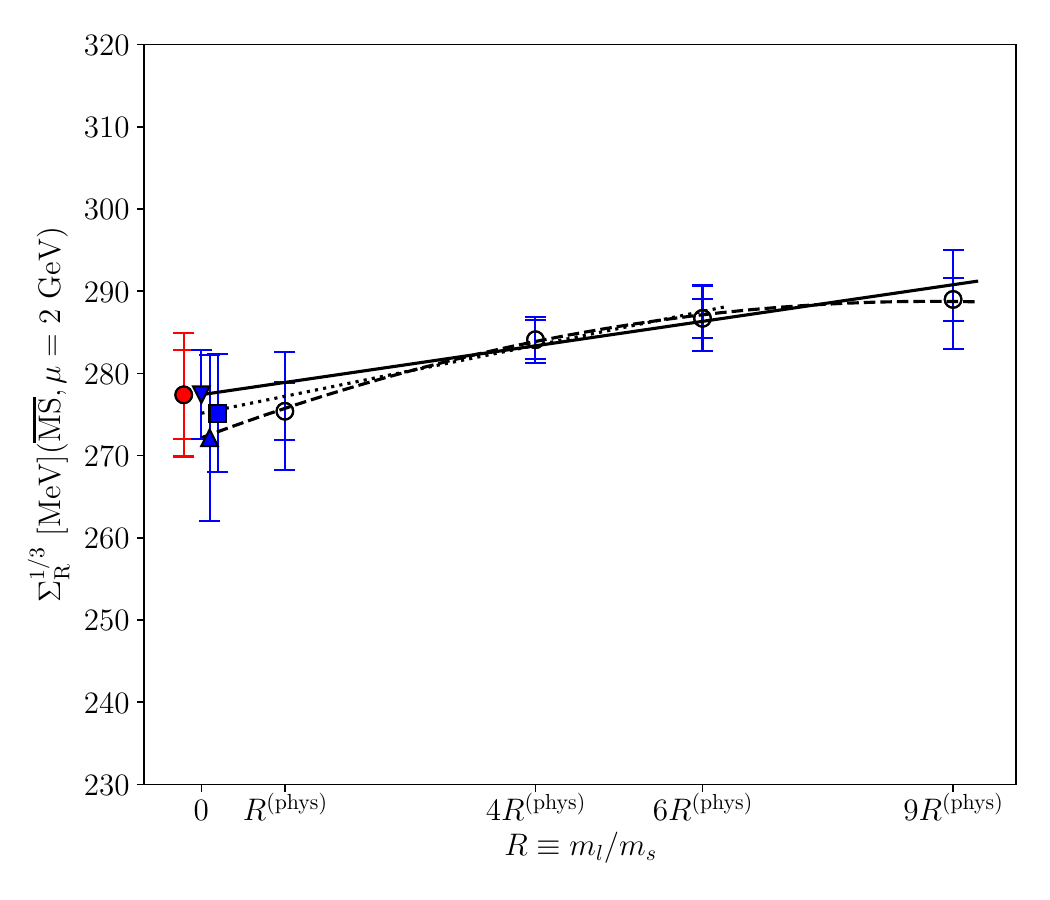}
	\caption{Left: continuum limit extrapolation of $\Sigma_{\mathrm{R}}^{1/3}$ for $m_l=4~m_l^{\phys}$. The solid line represents a linear fit in $a^2$ to the 3 finest lattice spacings, the dashed one the parabolic fit in $a^2$ including all points. The square point in $a=0$ is the result from the linear fit, the triangular point the one from the parabolic fit. Finally, the round point is the final estimation of $\Sigma_\R^{1/3}(R)$, with a double error bar referring to the statistical and to the sum of statistical and systematic uncertainties. Right: chiral limit extrapolation of $\Sigma_{\mathrm{R}}^{1/3}$ at fixed $m_s$ as a function of $R=m_l/m_s$. For each value of $R$, $\Sigma_\R^{1/3}(R)$ is reported with the usual double error bar. The dotted line represents the result of the best fit performed according to Eq.~\eqref{eq:fit_function_chiral_limit_linear_in_R} and including only the three smallest values of $R$. The straight line is the result from the same fit function, but including also $R=9~R^{\phys}$. The dashed line is the best fit result obtained by using a parabolic fit in $R$ in the whole range. The squared point at $R=0$ is the final result from the 3-point fit according to Eq.~\eqref{eq:fit_function_chiral_limit_linear_in_R}, the down-ward triangular point is the one from a linear 4-point fit and the up-ward triangular point is the result obtained from a parabolic 4-point fit. Finally, the round point represents the final value for the chiral condensate extracted from the mode number with a double error bar. Plots taken from Ref.~\cite{Bonanno:2023xkg}.}
	\label{fig:cc_mode_number_cont_chiral_limit}
\end{figure}

Once the determinations of the effective chiral condensate have been obtained as a function of the lattice spacing $a$ and of the light quark mass $m_l$, we adopt the following strategy to extract the chiral condensate. We first perform the continuum limit for each LCPs, i.e., for each value of $R$ (or $m_l$), by assuming standard $O(a^2)$ corrections:
\beq\label{eq:continuum_scaling_mode_number}
\Sigma_\R^{1/3}(a,R)=\Sigma_\R^{1/3}(R) + c_1(R)\,a^2+o(a^2).
\eeq
The continuum extrapolations for $m_l=4~m_l^{\phys}$ is shown in the left hand side of Fig.~\ref{fig:cc_mode_number_cont_chiral_limit}. It is clearly visible how the continuum scaling is pretty stable and well described by Eq.~(\ref{eq:continuum_scaling_mode_number}). The same holds for all other LCPs.

At this point, we computed the final value of the chiral condensate by performing a chiral limit according to the following relation~\cite{Giusti:2008vb}:
\beq\label{eq:fit_function_chiral_limit_linear_in_R}
\Sigma_\R^{1/3}(R)=\Sigma_\R^{1/3} + c_2\, R + o(R).
\eeq
The chiral limit is shown in the right hand side of Fig.~\ref{fig:cc_mode_number_cont_chiral_limit}. Again, Eq.~(\ref{eq:fit_function_chiral_limit_linear_in_R}) well describes the behaviour of our data. The final result for the chiral condensate from the mode number is:
\beq
\Sigma_\R^{1/3} = 277.4(5.4)_{\mathrm{stat}}(2.1)_{\mathrm{sys}}~\mathrm{MeV} \qquad \text{ (mode number)},
\eeq
where the central value is the result of the 4-point linear fit in $R$ and the systematic uncertainty takes into account the discrepancy between the central value and the result of the 4-point quadratic fit in $R$.
 
\section{$\Sigma$ from the pion mass and the topological susceptibility}
In this Section we briefly discuss other two methods to determine the chiral condensate that can be used to check our result from the mode number approach. The first one is based on the well-known relation from ChPT between the pion mass $M_\pi$ and the light quark mass, the so-called Gell--Mann-Oakes-Renner (GMOR) formula (for the case $m_u=m_d\equiv m_l$):
\beq\label{eq:gmor_formula}
M_\pi^2 = 2\frac{\Sigma}{F_\pi^2} m_l = 2 \left(\frac{\Sigma m_s}{F_\pi^2}\right) R.
\eeq
It is clear that, if one is able to compute the pion mass for each LCPs, i.e., for each value of $R$, then one can perform a linear fit in $R$ according to Eq.~(\ref{eq:gmor_formula}) and extract $\Sigma$ from the slope. In order to compute $\Sigma$ from the pion mass, we used the value of $m_s$ of Ref.~\cite{Davies:2009ih} while for $F_\pi$ we adopted the following result:
\beq\label{eq:pion_decay_constant_final}
F_\pi = 84.8(8.8)_{\mathrm{stat}}(6.1)_{\mathrm{sys}}~\mathrm{MeV}.
\eeq
obtained in Ref.~\cite{Bonanno:2023xkg}.

The pion mass can be computed by evaluating the large time behaviour of the correlator of the staggered interpolating operator of the physical pion and more details about the determination of $M_\pi$ can be found in the main paper~\cite{Bonanno:2023xkg}. 

\begin{figure}[!htb]
	\centering
	\includegraphics[scale=0.37]{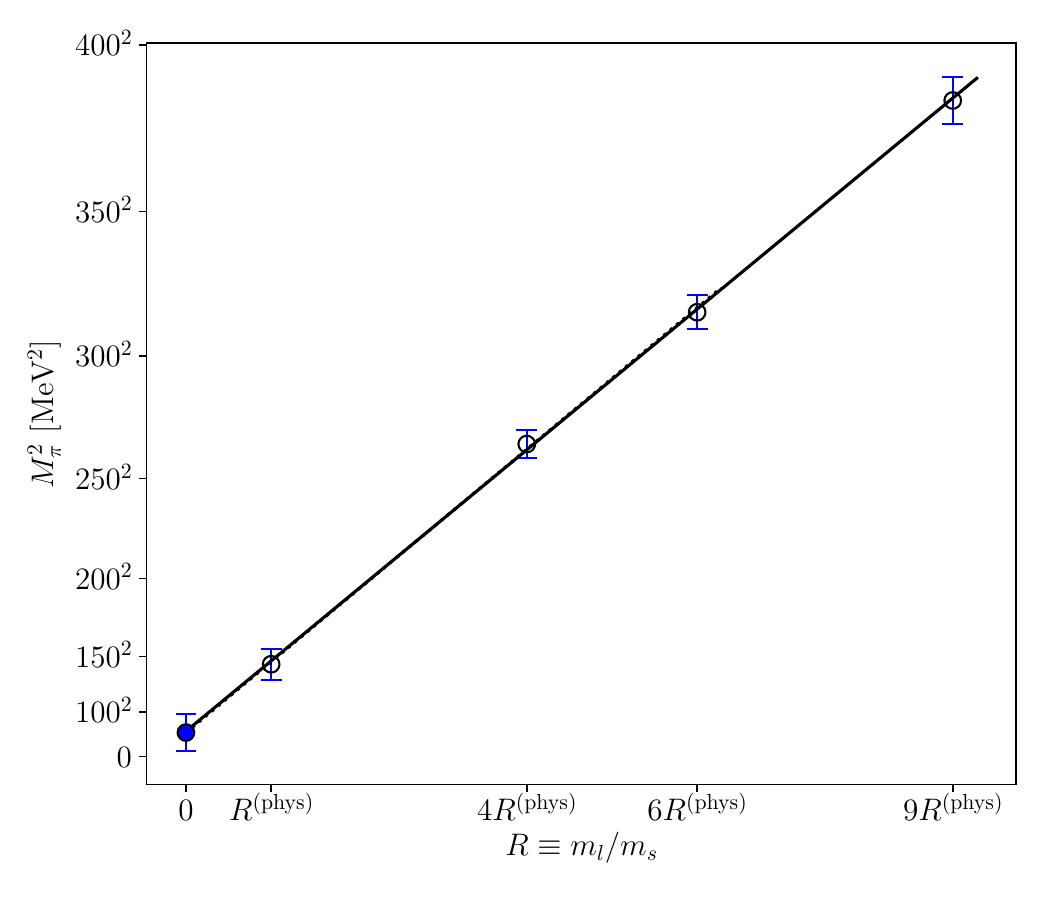}
	\includegraphics[scale=0.37]{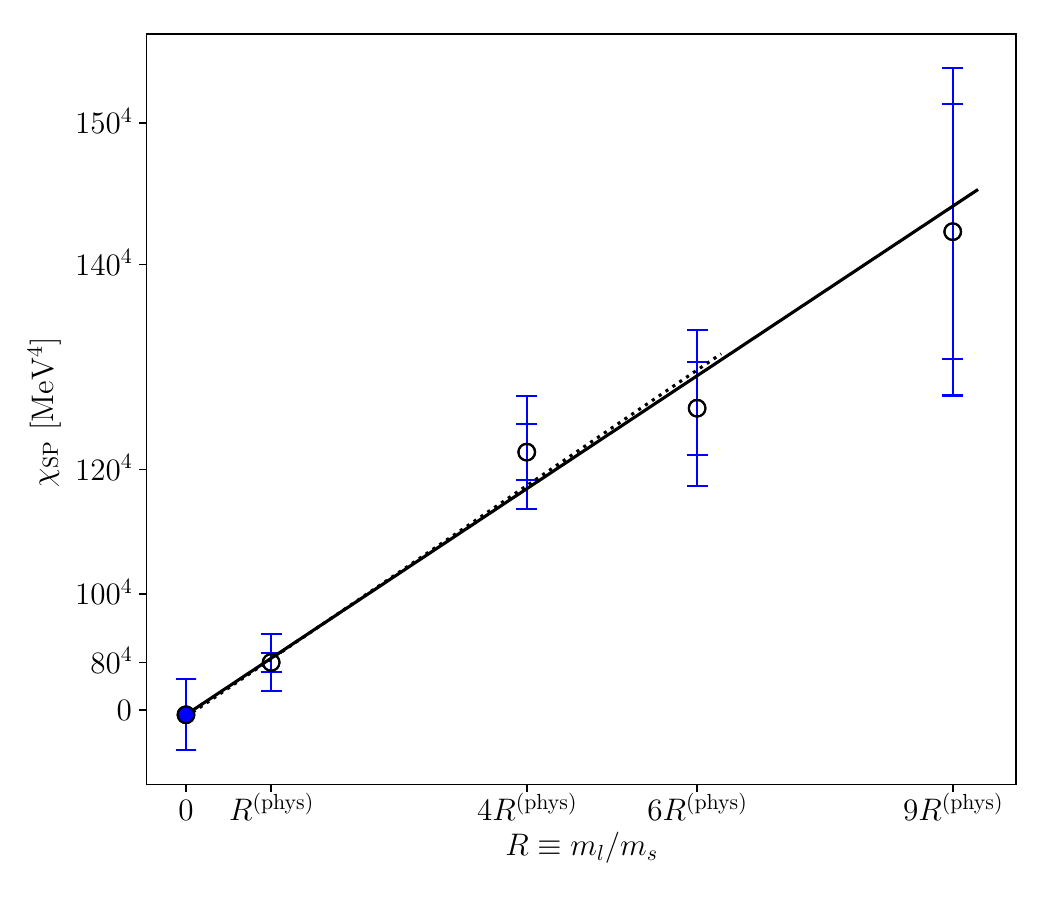}
	\caption{Left: extrapolation towards the chiral limit of $M_\pi^2$ at fixed $m_s$ as a function of $R=m_l/m_s$. The straight line is the result of a linear fit where the chiral limit  of $M_\pi^2$ is left as a free parameter. The dotted line, nearly indistinguishable from the solid one, is the best fit with respect to the same function but excluding the heavier pion mass. The filled point at $R=0$ is the chiral limit result including the whole range. In both cases, we find such fit parameter to be compatible with zero within errors. Right: extrapolation towards the chiral limit of the continuum results for the topological susceptibility $\chi$ at fixed $m_s$ as a function of $R$. The solid line is the result of a linear fit with the chiral limit left as a free parameter including the whole range. The dotted line refers to the best fit result obtained with the same function but excluding the point at $R=9R^{\phys}$. We find such fit parameter to be compatible with zero within errors. We adopt always the same convention for the double error bar. Plots taken from Ref.~\cite{Bonanno:2023xkg}.}
	\label{fig:mpi_chi_vs_R}
\end{figure}

We quote as final result for the chiral condensate from the pion mass the following one:
\beq\label{eq:final_result_sigma_from_mpi}
\Sigma_\R^{1/3} = 258.7(30.6)_{\mathrm{stat}}(0.01)_{\mathrm{sys}}~\mathrm{MeV} \qquad \text{ (pion mass)},
\eeq
where the systematic uncertainty, related to the discrepancy between the 3 and 4-point linear fit in $R$, turned out to be very small with respect to the statistical one. The chiral limit of $M_\pi^2$ is shown in the left hand side plot of Fig.~\ref{fig:mpi_chi_vs_R}.

The second approach is based on the quark mass dependence of the topological susceptibility $\chi$ from ChPT. We first recall the definition of $\chi$ as the two-point function of the topological charge density operator $q(x)$:
\beq\label{eq:q_definition}
q(x)=\frac{1}{32\pi^2} \varepsilon_{\mu\nu\rho\sigma} \Tr\left\{G_{\mu\nu}(x)G_{\rho\sigma}(x)\right\}
\eeq
\beq\label{eq:chi_definition}
\chi = \frac{1}{V}\int d^4x \langle q(x)q(0)\rangle =\frac{\langle Q^2\rangle}{V},\quad  Q=\int d^4x q(x)
\eeq
where $G_{\mu\nu}$ denotes the gauge field strength tensor and $Q$ is the topological charge.

From ChPT one has:
\beq\label{eq:ChPT_topsusc}
\chi = \frac{1}{2} \Sigma m_l=\frac{1}{2} \left[ \Sigma m_s \right] R,
\eeq
and, in a similar way to the pion mass approach, one can evaluate the functional dependence of $\chi$ on $R$ and extract $\Sigma$ from the slope of a linear best fit.

We use a fermionic definition based on Spectral Projectors \cite{Giusti:2008vb, Bonanno:2019xhg, Athenodorou:2022aay} to compute the topological susceptibility for each LCP. Here we report only the final result and refer the reader to Ref.~\cite{Bonanno:2023xkg} for the details: 
\beq
\Sigma_\R^{1/3} = 309.6(22.1)_{\mathrm{stat}}(0.2)_{\mathrm{sys}}~\mathrm{MeV} \qquad \text{ (topological susceptibility)},
\eeq
where the central value and the statistical error are the ones obtained from the 4-point linear fit, while the very small systematic one comes from the comparison between the results of the 3-point and the 4-point fits.

\section{Discussion of the results and conclusions}
We have addressed the computation of the flavour SU(2) chiral condensate from a staggered discretization of $N_f=2+1$ QCD. The main novelty is the application of the Giusti--L\"uscher method (introduced in Ref.~\cite{Giusti:2008vb} for Wilson fermions) based on the Banks--Casher relation to the staggered case. We also checked our result from the mode number approach with other ones based on the relation, coming from ChPT, of the chiral condensate with the pion mass and the topological susceptibility.

We report all our determinations of the chiral condensate:
\beq
\Sigma_\R^{1/3} &= 277.4(5.4)_{\mathrm{stat}}(2.1)_{\mathrm{sys}}~\mathrm{MeV}  \qquad &\text{ (from $\braket{\nu}$)},\\
\Sigma_\R^{1/3} &= 258.7(30.6)_{\mathrm{stat}}(0.01)_{\mathrm{sys}}~\mathrm{MeV} \qquad &\text{ (from $M_\pi$)},\\
\Sigma_\R^{1/3} &= 309.6(22.1)_{\mathrm{stat}}(0.2)_{\mathrm{sys}}~\mathrm{MeV} \qquad &\text{ (from $\chi$)}.
\eeq
They are all compatible among themselves and it is worth noticing that the one based on the mode number has a much better precision.

At this point, it is reasonable to perform a global fit of these quantities as function of the $R\equiv m_l/m_s$ constraining them to give the same result for $\Sigma_\R$. Here we just claim that such a kind of fit has a reasonable chi-squared and gives the following result, which we take as our final determination for $\Sigma_\R$:
\beq\label{eq:final_res}
\Sigma_\R^{1/3} = 265.7(4.2)_{\mathrm{stat}}(0.5)_{\mathrm{sys}}~\mathrm{MeV}, \qquad \text{ (global fit)}.
\eeq
The central value and the statistical uncertainty come from the 4-point linear best fit in $R$ where the chiral limits of $M_\pi$ and $\chi$ are left as free parameters. The systematic uncertainty, which is small if compared to the statistical one, takes into account the discrepancy between the previous fit and the one made with 4 points where $M_\pi(R=0)$ and $\chi(R=0)$ are fixed to zero. We adopted this criterium because the difference between 3 and 4-point fits was always negligible.

Since the three different computations of $\Sigma_\R^{1/3}$ come from the same ensembles of gauge configurations, they are surely correlated. However, we expect that these correlations are not so significant because the observables involved are very different and affected by systematics of different orders of magnitude. Furthermore, since the estimation of the chiral condensate coming from the mode number is much more precise than the other ones, we expect the final result being mainly driven from it. As final comment, we notice that the final result is perfectly in agreement with the world-average value of $\Sigma_\R^{1/3}$, obtained from $N_f=2+1$ QCD, quoted in the latest FLAG review: $\Sigma_{\mathrm{FLAG}}^{1/3} = 272(5)$ MeV~\cite{FlavourLatticeAveragingGroupFLAG:2021npn}.

\section*{Acknowledgements}
We thank L.~Giusti for useful discussions. The work of C.~Bonanno is supported by the Spanish Research Agency (Agencia Estatal de Investigación) through the grant IFT Centro de Excelencia Severe Ochoa CEX2020- 001007-S and, partially, by grant PID2021-127526NB-I00, both funded by MCIN/AEI/ 10.13039/ 501100011033. C.~Bonanno also acknowledges support from the project H2020-MSCAITN-2018-813942 (EuroPLEx) and the EU Horizon 2020 research and innovation programme, STRONG-2020 project, under grant agreement No 824093. Numerical simulations have been performed on the \texttt{MARCONI} and \texttt{MARCONI100} machines at CINECA, based on the Project IscrB ChQCDSSP and on the agreement between INFN and CINECA (under projects INF22\_npqcd, INF23\_npqcd).

\providecommand{\href}[2]{#2}\begingroup\raggedright\endgroup

\end{document}